\documentclass{aa}
\usepackage{natbib}
\usepackage{graphicx}
\usepackage{amssymb}
\usepackage{multirow}
\usepackage[varg]{txfonts}

\bibpunct{(}{)}{;}{a}{}{,}

\begin{document}

\title{The stellar metallicity gradients  in galaxy discs\\ in a cosmological scenario}

\author{Patricia B.~Tissera\inst{1,2} \and Rubens E.~G.~Machado\inst{1} \and Patricia S\'anchez-Bl\'azquez\inst{3} \and Susana E.~Pedrosa\inst{4} \and Sebasti\'an F.~S\'anchez\inst{5} \and Owain N.~Snaith\inst{6} \and Jos\'e M.~Vilchez\inst{7}}
\offprints{P. B. Tissera}
\institute{Departamento de Ciencias Fisicas, Universidad Andres Bello,
Av. Republica 220, Santiago, Chile.\\
\email{patricia.tissera@unab.cl}
\and Millennium Institute of Astrophysics, Chile.
\and Institute of Astronomy, Av. Vicu\~na Mackena, Pontificia Universidad Cat\'olica de Chile.
\and Instituto de Astronom\'{\i}a y F\'{\i}sica del Espacio,
CONICET-UBA, Casilla de Correos 67, Suc. 28,  C1428ZAA, Ciudad
Aut\'onoma de Buenos Aires, Argentina.
\and Instituto de Astronom\'ia, Universidad Nacional Aut\'onoma de M\'exico, A.P. 70-264, 04510 M\'exico D.F., Mexico
\and School of Physics, Korea Institute for Advanced Study, 85 Hoegiro,
Dongdaemun-gu, Seoul 02455, Republic of Korea 
\and Instituto de Astrof\'isica de Andaluc\'ia (CSIC), Glorieta de la Astronom\'ia s/n, E-18008 Granada, Spain.
}

\date{Received / Accepted}

\abstract
{The stellar metallicity gradients of disc galaxies provide information on the disc assembly, star formation processes and chemical evolution. They also might store information on dynamical processes which could affect the distribution of chemical elements in the gas-phase and the stellar components. Understanding their joint effects within a hierarchical clustering scenario is of paramount importance. }
{We studied the stellar metallicity gradients of simulated discs in a cosmological simulation. We explored the dependence of the stellar metallicity gradients on stellar age and the size and mass of the stellar discs.}
{We used a catalogue of galaxies with disc components selected from a cosmological hydrodynamical simulation performed including a physically-motivated Supernova feedback and chemical evolution. Disc components were defined based on  angular momentum and binding energy criteria. The metallicity profiles were estimated for stars with different ages. We confront our numerical findings with results from  the {\small CALIFA} Survey.}
{The simulated stellar discs are found to have metallicity profiles with slopes in global agreement with observations. Low stellar-mass galaxies tend to have a larger variety of metallicity slopes. When  normalized by the half-mass radius, the stellar metallicity gradients do not show any dependence and the dispersion increases significantly, regardless of the galaxy mass. Galaxies with stellar masses around $10^{10}$M$_{\odot}$ show  steeper negative metallicity gradients. The stellar metallicity gradients  correlate with the half-mass radius. However, the correlation signal is not present when they are normalized by the half-mass radius. Stellar discs with positive age gradients are detected to have negative and positive metallicity gradients, depending on the relative importance of the recent star formation activity in the central regions.}
{ Our results suggest that the inside-out formation is the main process reponsible for  the metallicity and age profiles. The large dispersions in the metallicity gradients as a function of stellar mass could be ascribed to the effects of dynamical processes such as mergers/interactions and/or migration as well as those regulating the conversion of gas into stars. The fingerprints of the inside-out formation seem better preserved by the stellar  metallicity gradients as a function of the half-mass radius. }

\titlerunning{Stellar metallicity gradients}
\authorrunning{Tissera et al.}
\maketitle

\section{Introduction}

Chemical patterns  such as  the mass-metallicity relation, the age-metallicity relation and  the metallicity gradients store relevant information on the history of galaxy assembly. Among them, observational results show that slope of the gas-phase metallicity gradients  correlates with stellar mass if expressed in ${\rm dex~kpc^{-1}}$ but that this correlation is erased if the metallicity gradients are renormalized by adopting a characteristic scale-length for the disc component \citep[e.g.][]{garnett1997,sanchez2014,ho2015,sanchezmenguiano2016}. Similar results are found for discs formed in hydrodynamical cosmological simulations \citep{tissera2015} and by using analytical models \citep{prantzos2000,ho2015}. Both observational and theoretical results favour the existence of metallicity gradients the  origin of which would be linked to the joint action of gas outflows and inflows which regulate the star formation activity and the chemical abundances \citep[e.g.][]{chiappini2001,molla2005,gibson2013,mott2013}. However, an important fraction of the synthesized chemical elements is locked into stellar populations (SPs) as the gas is transformed into stars. Hence, the chemical abundances of the SPs provide complementary information to understand the chemical loop between the gas-phase medium and the SPs within a context of galaxy formation \citep[e.g.][]{calura2012,tissera2012}.

Most observational studies are devoted to the gas-phase abundances in discs traced by HII regions since the study of the chemical properties of the underlying SPs is more difficult. Estimations for the Milky Way  and some nearby galaxies suggested the existence of  negative metallicity slopes \citep[e.g.][]{freeman,carraro2007,hughes1994,kudritzki2008}. The metallicity properties of the  SPs in galaxies outside the Local Group have been studied by using colour distributions. In these works, the colour gradients are interpreted as the result of the contribution of younger and less enriched stars  to the outer parts \citep[e.g.][]{dejong1996,macarthur2004,munoz2007}. 2D observations of nearby galaxies with detailed information on the age and metallicities of the stellar populations are being gathered. As a result, more robust properties of the SPs in nearby galaxies are available \citep[e.g.][]{rosalesortega2010}. In particular, \citet{sanchezb2014} reported normalized metallicity gradients of SPs in the disc components for a set of face-on Spirals  in  the CALIFA survey. Similar to the results reported for the gas-phase normalized  metallicity gradients \citep{sanchez2014}, the SP normalized  metallicity slopes were not found to correlate with the global properties of galaxies such as stellar mass or luminosity. A weak trend for the old stellar population ($> 6$~Gyr) to show steeper metallicity gradients than younger ones ($< 2 $~Gyr) was reported.  The metallicity and age gradients of galaxies with a variety of morphology are analysed by \citet{gonzalezdelgado2015}. These authors also find a dependence of both gradients on stellar mass (or the potential well). In particular they found that Spiral galaxies have negative metallicity and age profiles in agreement with an inside-out formation. For disc-dominated galaxies, they identified the steeper negative metallicity slopes to take place for intermediate stellar mass galaxies \citep{molla2005}.

The metallicity gradients of the SPs are the result of the combination of stars formed at different epochs from gas in  the interstellar medium (ISM) which have different physical and chemical properties. Disentangling their contributions is a complex task, albeit crucial to build up a comprehensive understanding of  how star formation, chemical production and mixing take place  in galaxies. Mergers and interactions are known to be efficient mechanisms to redistribute angular momentum and as a consequence they could affect the metallicity distribution of the gas and stars. The action of migration has been also claimed to be important in shaping the metallicity gradients, flattening the abundance profiles of the SPs, principally of the old ones \citep[e.g.][]{friedli1994, roskar2008, dimatteo2013}. However, more realistic simulations where the  vertical velocity of the stars is taken into account, suggest that strong migration might be prevented \citep{minchev2014}.

In the current cosmological paradigm, galaxy formation is a complex process which involves the action of different physical mechanisms acting with different temporal and spatial scales. In this context, hydrodynamical simulations which include chemical evolution are powerful tools to study the relation between chemical patterns and  the history of galaxy assembly \citep[e.g.][]{mosconi2001,scan06,tissera2012,volgel2014,schaye2015}. Similarly, chemodynamical models have extensively analysed the formation of metallicity gradients in disc galaxies \citep[e.g.][]{chiap1997,molla1997,ho2015}.  Within a hierarchical clustering scenario, different mechanisms play a role in the assembly history of galaxies, however most numerical works suggest the inside-out formation for the disc components \citep{scan09,brook2012,few2012,minchev2014}.  Particularly, \citet{pilkington2012} analysed the metallicity gradients of the young stellar populations as tracers of the gas-phase metallicity,  finding that the efficiency of star formation as a function of radius was a main factor determining the metallicity gradients. And as expected, different Supernova (SN) feedback models produced different results \citep{gibson2013,snaith2013}. \citet{tissera2015} analysed the metallicity gradients of the gas-phase of  the same set of simulated galaxies analyzed in our work. They detected a trend for the gas-phase metallicity slopes to correlate with stellar mass. These authors found a relationship between the efficiency of star formation and the slope of the  gas-phase metallicity profile in agreement with the observations of \citet{stott2014}. According to detailed numerical simulations of prepared merger events, these relationships could be the result of the action of  violent events in the regulation of the star formation and chemical mixing \citep[][and Sillero in preparation]{tissera2015}. 

In this work, we focused on the study of  the metallicity gradients of the SPs in the disc components of the same simulated galaxies whose gas-phase components were analysed by \citet{tissera2015} in the redshift range [0,2). We use the 12~+log~ O/H abundance ratios  and explore correlations with other physical parameters such as the stellar mass and  the half-mass radius. We also extended the analysis of   the metallicity gradients for SPs with different ages. We performed this analysis with a two-fold goal: searching for clues for galaxy formation and setting contraints on our galaxy formation model.

The paper is organized as follows. Section 2 explains the main characteristics of the numerical models and the galaxy catalogue. Section 3 is dedicated to the analysis of the SP metallicity gradients. Conclusions summarizes the main findings.

\section{Numerical Experiments}
\label{sec:simus}

We analysed the stellar disc components of simulated galaxies selected from a cosmological simulation performed with a version of the  code {\small GADGET-3}, an update of {\small GADGET-2 } \citep{springel2005}, optimized for massive parallel simulations of highly heterogeneous systems. This version includes treatments for metal-dependent radiative cooling, stochastic star formation (SF), chemical enrichment, and the multiphase model for the ISM and the SN feedback scheme of \citet{scan05,scan06}. The SN feedback model is able to successfully trigger galactic mass-loaded winds without introducing mass-scale  parameters.

Our code considered energy feedback by Type II (SNII) and Type Ia (SNIa) Supernovae. These processes  are grafted into a multiphase model for the ISM described in detail by  \citet{scan06}. Briefly, the energy released by each SN event   is distributed between the cold and the hot phases of the  simulated ISM. The adopted ISM  multiphase model  allows the coexistence of diffuse and denses gas phases. In this model, each gas particle defines the cold and hot phases by using a local entropy criteria. This scheme allows particles to decouple hydrodynamically from particular low-entropy ones if they are not part of a shock front. Each cold gas particle  stores the SN energy received from nearby SN events until it fullfils the conditions to join their local hot phase. It is worth mentioning that this SN feedback scheme does not include parameters that  depend on the global properties of the given galaxy (e.g. the total mass, size) thus making it suitable for cosmological simulations, where systems with different masses form in a complex way. The SN energy is equally distributed between the cold and hot gas-phases, surrounding a stellar source.

As mentioned in the Introduction, we analysed the SPs of the simulated disc components whose gas-phase medium was studied by \citet{tissera2015}. They were selected from the   so-called S230D simulation, whose  initial conditions are consistent with the concordance model with $\Omega_{\Lambda}=0.7$, $\Omega_{\rm m}=0.3$, $\Omega_{b}=0.04$, a normalization of the power spectrum of $\sigma_{8}=0.9$ and $H_{0}= 100 h \ {\rm km} \ {\rm s}^{-1}\ {\rm Mpc}^{-1}$, with $h=0.7$. The simulated volume represents a box of $14$ Mpc comovil side, resolved  with $2 \times 230^3$ initial particles, achieving a mass resolution of $5.9\times 10^{6}{\rm h^{-1}\ M_{\odot}}$ and $9.1\times 10^{5}{\rm  h^{-1}\  M_{\odot}}$ for the dark matter and initial gas particles, respectively. The gravitational softening is $0.7$~kpc.   We acknowledge the fact that the initial condition represents a small volume of the Universe. It  was chosen to represent a typical region, with no large structure nearby. \citet{derossi2013}  verified that the growth of the haloes was well-reproduced by confronting the halo mass function with those obtained from the Millennium simulation  \citep{fakhouri2010}.

Simulation S230D used a high gas density threshold for the star formation and an  energy per SN event of $0.7 \times 10^{51 }$~erg. This combination diminishes the over transformation of gas into stars at very high redshift. Simulation S230D injects   $80$ per cent of the new synthesized chemical elements into the cold phase (the remaining is pumped directly into the hot phase).  Our code also includes the chemical evolution model developed by \citet{mosconi2001}. SNII and SNIa events contribute with chemical elements estimated by adopting the chemical yields of \citet{WW95} and \citet{iwamoto1999}, respectively. Note that we follow twelve individual chemical elements (such as $^{16}$O, $^{56}$Fe) and that SNII yields are metallicity dependent \citep{mosconi2001}.

\begin{figure*}
\includegraphics[width=0.8\textwidth]{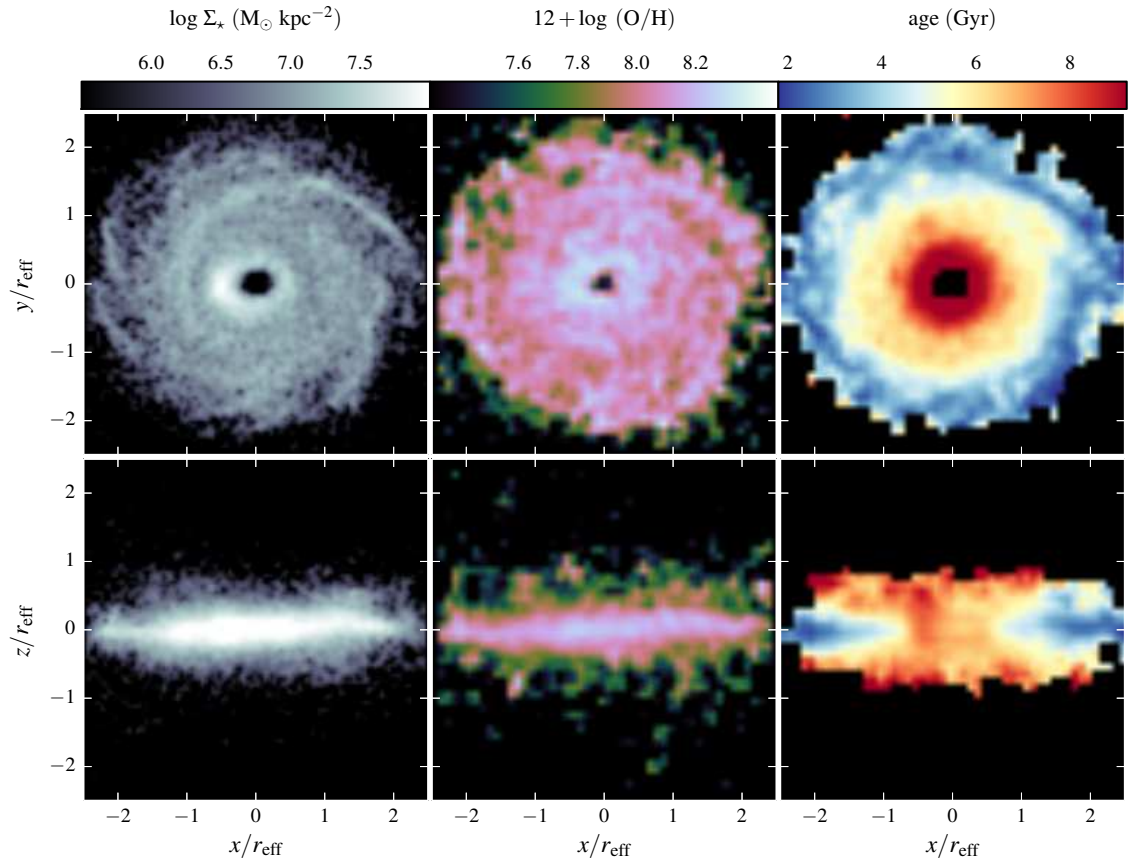}
\centering
\caption[]{Surface mass density, oxygen abundance,
  and  ages for the stellar populations in a  simulated discs shown
  for illustration purposes (only stars with $\epsilon < 0.5$ have
  been included) . Each frame is $5r_{\rm eff}$ wide, and the half-mass radius of this stellar disc is $r_{\rm eff} = 11$ kpc.}
\label{fig:maps}
\end{figure*}

\subsection{The simulated galaxy sample}

The simulated galaxy sample was constructed  by using a Friends-of-Friends algorithm to identify the virialized structures. Then the  substructures were selected by  using the SUBFIND scheme \citep{springel2001}. In order to identify the disc components,  we used the angular momentum content of each particle  defined as $\epsilon = J_{\rm z} /J_{\rm z,max}(E)$, where $J_{\rm z}$ is the angular momentum component in the direction of the total angular momentum, and $J_{\rm z,max}(E)$ is the maximum $J_{\rm z}$ over all particles at a given binding energy, $E$. The details are explained in \citet{tissera2012} and \citet{pedrosa2015} We considered star particles with $\epsilon$ higher than $0.5$ to be part of the stellar disc. The star particles  that did not satisfy this requirement were taken to belong to the spheroidal component and are not analysed in this paper. We do not distinguish between thin and thick discs. They are both included in the analysed stellar disc components. To diminish resolution issues, we restricted our study to systems with a number of star particles larger than $2000$ in  the disc components. With this criterium, 37 galaxies were selected at $z\sim 0$. 

For illustration purposes, in Fig.~\ref{fig:maps} we show face-on and edge-on projected distributions of the stellar mass, the oxygen abundances and the stellar ages for one of the simulated disc components. The bulge has been extracted by applying the procedure described above.  From this figure, we can see that existence of spiral arms which are populated mainly by young stars.  In this example, there are clear age and metallicity profiles, although there are also local inhomogenities, revealing a complex structure which varies from system to system. 

The simulated galaxies analysed in this paper were also studied by \citet{pedrosa2015}. These authors focused on the angular momentum content of the discs and bulges components, finding that the simulated galaxies are able to reproduce  observed trends reported by \citet{romanowsky2012}. Hence, we expect the sizes of the simulated galaxies to be in agreement with observations. \citet{tissera2015} studied the gas-phase metallicity gradients of the same galaxies we analyse in this paper, in relation to the specific star formation of the galaxies. These authors reported that these galaxies reproduced observational trends at $z\sim 0$. In agreement with observations, \citet{tissera2015} showed that simulated gas-phase metallicity slopes of these low stellar-mass galaxies  exhibit a larger variety. As a function of redshift these authors found an increasing fraction of gas-phase metallicity profiles with  negative slopes ($<-0.1$dex kpc$^{-1}$ in galaxies with $M < 10^{10.5}$M$\odot$). More massive galaxies showed larger frequency of positive metallicity slopes for higher redshifts. 

We  use the CALIFA sample of \citet{sanchezb2014} to confront the simulated trends at $z\sim 0$ since it is a self-consistent sample and the stellar mass range covered is similar to our simulated mass range. For the purpose of comparison, we use the mass-weighted metallicity gradients. We note that the observed half-mass radii are defined from the surface luminosity distributions while we use the half-mass radius of the stellar disc components. This should be bore in mind when comparing the normalized gradients.

\section{Metallicity and  age gradients of the SPs }
\label{sec:metgrad}

\begin{figure}
\includegraphics[width=\columnwidth]{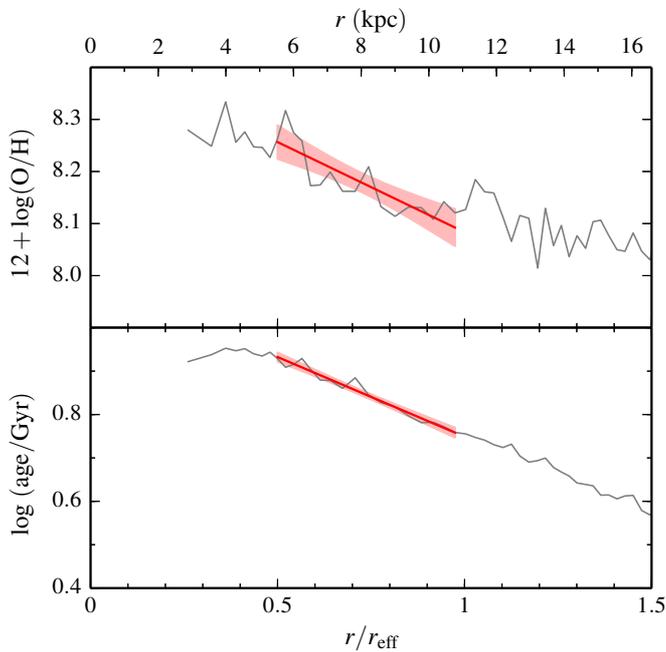}
\centering
\caption[]{Metallicity and  age profiles for the SPs in the stellar disc 
component of 
  the galaxy shown in  Fig.~\ref{fig:maps} (thin grey
  lines).  The thick red lines
  are the linear fits within the region $0.5 < r/r_{\rm eff} < 1$.
The shaded areas represent the 95 per cent  confidence bands of the fits.}
\label{fig:profiles}
\end{figure}

We estimated the radial abundance gradients and age profiles of the disc SP.  We performed linear regression fits to the profiles in the range $[0.5, 1] r_{\rm eff} $ where $r_{\rm eff} $ is the radius which enclosed half the stellar mass of the disc components. We find that this radial region provides stable and good fits for our profiles and it is within the radial range chosen by \citet{sanchezb2014} for the analysis of the SP gradients in a set of face-on spiral galaxies of the CALIFA survey which will be used to compare with our simulated data. This is also the spatial range adopted to fit the gas-phase metallicity gradients  by \citet{tissera2015}. The linear regressions are performed by applying a bootstrap technique.  In Fig.~\ref{fig:profiles} we show the metallicity  and age profiles (upper and lower panels, respectively) for the same galaxy depicted in Fig. ~\ref{fig:maps}. The goodness of the fittings are shown by the shaded areas (95 per cent confidence bands). 

\subsection{Metallicity gradients}

\begin{figure}
\includegraphics[width=\columnwidth]{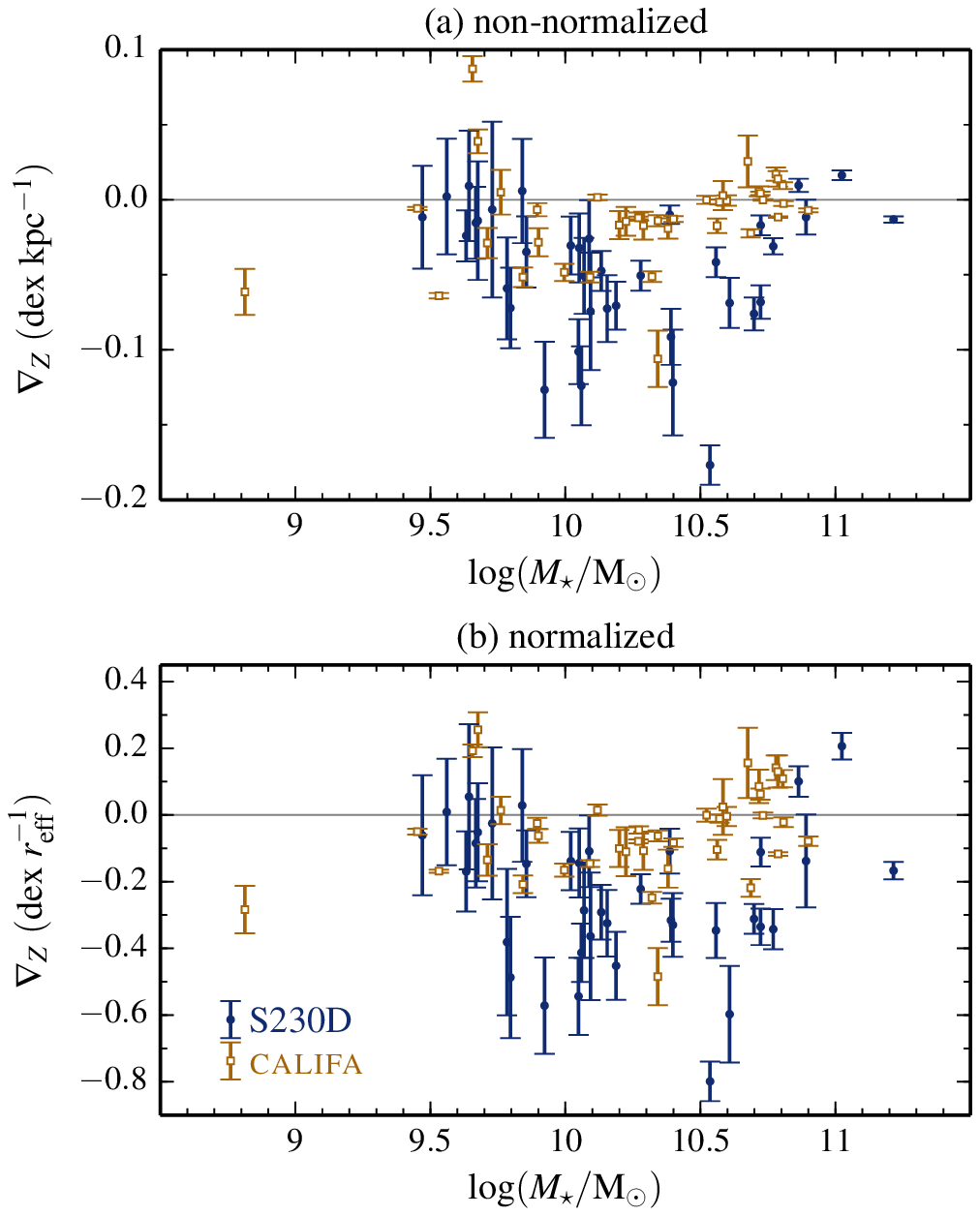}
\centering
\caption[]{Slopes of the non-normalized (upper panel)  and normalized (lower panel) metallicity gradients of the stellar populations in the
  disc components as a function of total  stellar mass of the
  simulated galaxies (solid, blue circles). 
  The error bars correspond to a
bootstrap errors of the linear regression fits. Observations from the CALIFA survey  by \citet[][open, yellow
  squares]{sanchezb2014} have
  been included for comparison. }
\label{fig:metM}
\end{figure}

Figure ~\ref{fig:metM} displays the metallicity slopes in units of dex kpc$^{-1}$ (upper panel) and the renormalized ones by  using $r_{\rm eff} $ (lower panel) as a function of the stellar mass of the galaxies.  For comparison, we also include the observational estimations reported by \citet{sanchezb2014}. 

In order to quantify a possible dependence on the stellar mass, we define two subsamples by adopting the stellar mass limit,  $M < 10^{10.5}$M$\odot$, similarly to the analysis carried out by \citet{tissera2015} for the gas-phase metallicity profiles. From  Fig.~\ref{fig:metM}  and Table~\ref{table:1}, we can see that the non-normalized metallicity slopes show a very weak dependence on stellar mass. Galaxies in the low-stellar mass subsample have  a median metallicity slope of  $-0.04 \pm 0.01$  ${\rm dex~{ kpc}^{-1}}$ while those in the high-stellar mass subsample have  $-0.03 \pm 0.02$  ${\rm dex~{ kpc}^{-1}}$. The median slopes and errors have been estimated by applying a bootstrap resampling technique. In Table~\ref{table:1} we also include the standard dispersions. The most significant variation with stellar mass is the increase of the dispersion of the metallicity slopes measured in  the low stellar-mass galaxies.

Similar estimations for the normalized SP metallicity gradients have been carried (${\rm dex~{\it r}_{eff}^{-1}}$). As can be seen from the lower panel of from  Fig.~\ref{fig:metM},  there is no dependence on stellar mass and the dispersion increases significantly for all galaxies  (Table~\ref{table:1}). It is also interesting to note that there are galaxies with normalized metallicity profiles with positive slopes at the two extreme of the stellar-mass range. And a similar behaviour is also present for the CALIFA slopes. In the case of those at the lower mass  end, the fittings have a larger dispersion. The Spearman test of correlation yields no  signal of correlation for both distributions ($r=-0.11, p = 0.51$ for the non-normalized relation and ($r=-0.19, p= 0.27 $ for the normalized relation). We note that the gas-phase metallicity gradients of these simulated galaxies show clear trends with stellar mass as reported by \citet{tissera2015}. The weakness of the signal could be due to the action of  dynamical processes or different star formation efficiencies. The SN feedback adopted for this simulation is able to regulate the transformation of gas into stars naturally in different potential wells \citep{derossi2010}. Interestly,  the simulated discs are able to reproduce a trend to have the steeper negative metallicity gradients for masses around $10^{10} M\odot$ as shown by \citet{sanchezb2014} and \citet{gonzalezdelgado2015} \citep[see also][for similar results by using analytical models.]{molla2005}. \citet{derossi2013} showed that there is a change in the effects of the SN feedback around this stellar mass due to combined effects of the SN strengths and the potential well of the systems. These changes affect the gas cooling time,  star formation activity and the amount of metals ejected in galactic winds. We speculate that these processes might also leave an imprint in the chemical gradients. A large stastistical sample is needed to confirm these trends.

\begin{figure}
\includegraphics[width=\columnwidth]{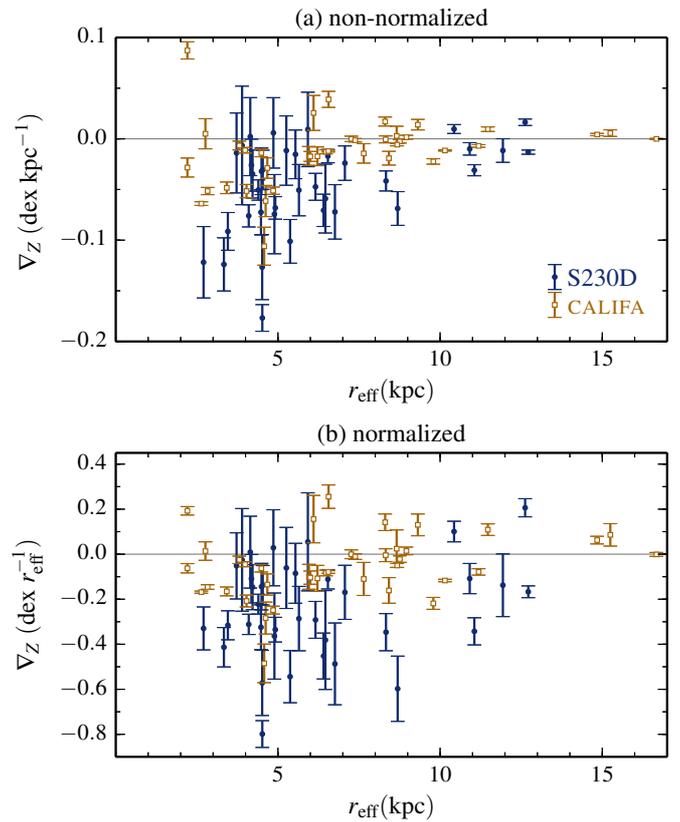}
\centering
\caption[]{Slopes of the non-normalized (upper panels)  and normalized
  (lower panels) metallicity gradients of the SPs in the
  disc components as a function of $r_{\rm eff}$ (solid, blue circles). The error bars correspond to the 
bootstrap errors of the linear regression fits. }
\label{fig:metR2}
\end{figure}

\citet[][see also \citet{ho2015}]{prantzos2000} reported a correlation of the metallicity gradients of the gas-phase with the $r_{\rm  eff}$ which got erased if the gradients were normalized by a characteristic scale-length. In order to test the existence of such a correlation for the simulated SPs, we show in Fig.~\ref{fig:metR2} the relation for the simulated discs. In this case, we find  a clearer signal of correlation for the non-normalized gradients: $r \sim 0.38, p\sim 0.02$. The correlation signal significantly decreases for  the normalized gradients: $r \sim 0.031, p\sim  0.86$.

\begin{figure}
\includegraphics[width=\columnwidth]{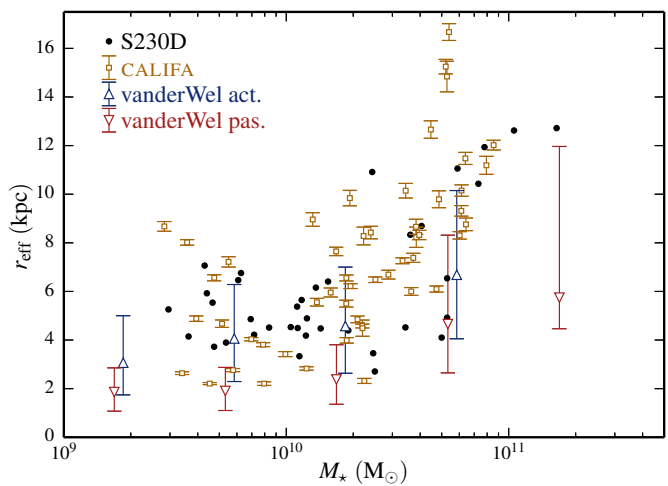}
\centering
\caption[]{ Mean $r_{\rm eff}$ of the stellar disc components as a function of total  stellar mass of the
  simulated galaxies (solid, blue circles) for the observational
  estimations by \citet{wel2014}  for early-type (red, inverted triangles) and
  late-type (blue triangles) galaxias and for the observations of the
  disc galaxies in the CALIFA survey by \citet[][open, yellow
  squares]{sanchezb2014}. }
\label{fig:sizeM}
\end{figure}

Since the characteristic radius  $r_{\rm  eff}$ seems to  play an important role, it is important to show that the simulated $r_{\rm  eff}$ are in agreement with observations. In Fig.~\ref{fig:sizeM} we display them a function of stellar mass  together  with recent observations by  \citet{wel2014} and \citet{sanchezb2014}. As it can be seen from this plot,  the mean values are consistent with observations by \citet{wel2014}  within a standard deviation for late-type galaxies as expected since most of our galaxies have active star formation \citep{tissera2015}. However, at the more massive end, our simulated systems show larger values than the reported averages. {The simulated  $r_{\rm  eff}$  are in very good agreement with  those estimated from the CALIFA survey}\footnote{ We also fitted exponential discs to the mass surface density and calculated the $r_{\rm eff}$  from the disc scale-lengths. These values agreed with those estimated directly from the mass distributions.}.

\subsection{Age gradients}

The stellar age profiles store information on the formation history of the stellar disc.  In an inside-out disc formation scenario negative age gradients are expected since the central regions formed first and have higher density gas to feed the star formation activity. Although this might be the case on global terms, mergers and interactions as well as secular evolution could contribute to flatten or to build positive age and/or metallicity profiles as shown in previous works \citep[e.g.][]{rupke2010,perez2011}. Migration could also cause old stars to populate outer regions, flattening the age and metallicity distributions. Minor mergers might also add old stars to the outer parts of the discs \citep[][]{abadi2003,tissera2012}.

\begin{figure*}
\includegraphics[width=\columnwidth]{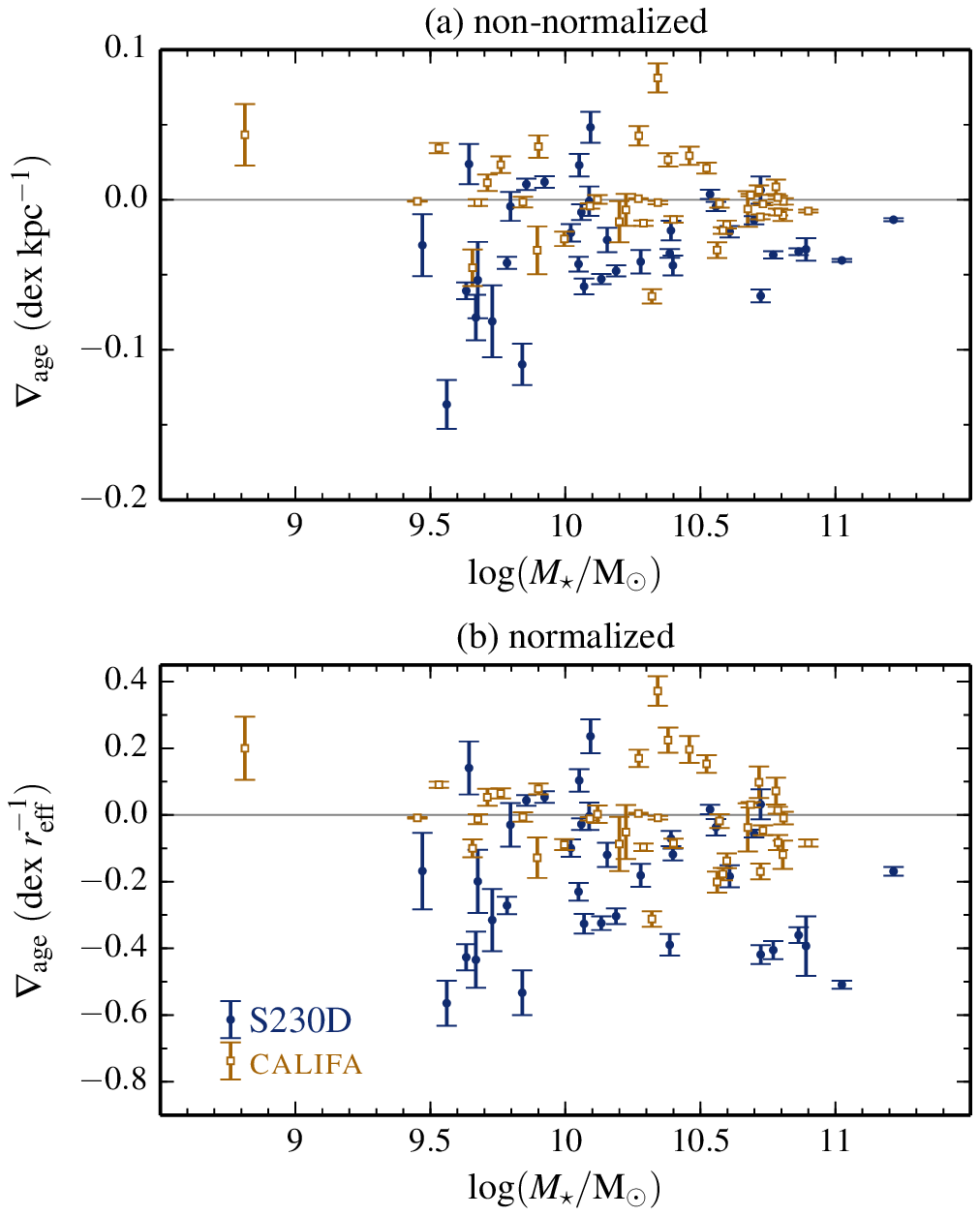}
\includegraphics[width=\columnwidth]{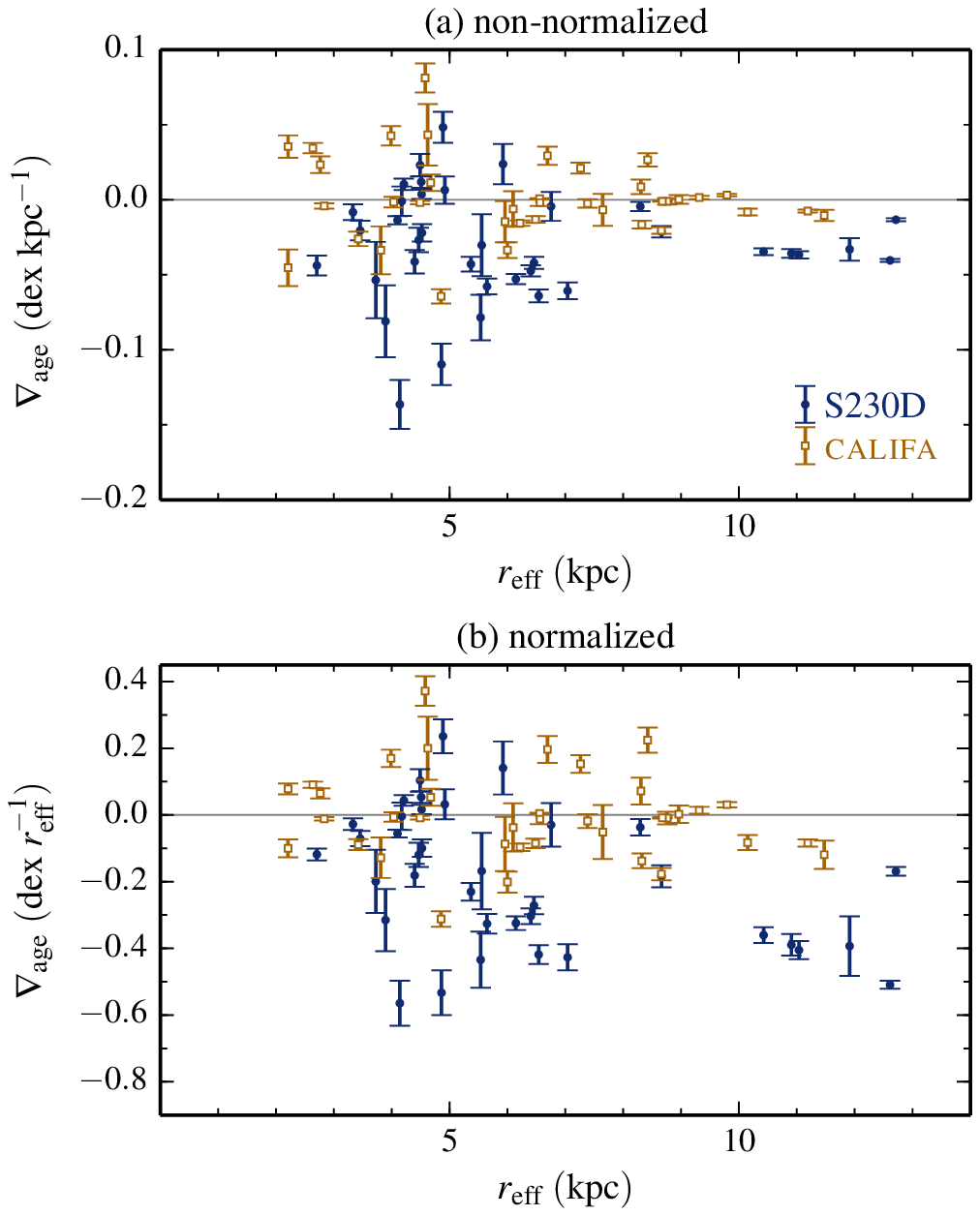}
\centering
\caption[]{Slopes of non-normalized (upper panel) and normalized (lower panel) age gradients of the stellar
  populations as a function of the stellar mass (left panles) and
  $r_{\rm eff}$ (right panels) the of the simulated  galaxies.
  Similarly, observations from the CALIFA survey  \citet[][open, yellow
  squares]{sanchezb2014} have
  been included for comparison.
}
\label{fig:ageM}
\end{figure*}

We  fit  linear regressions to the age profiles following the same procedure as in the case of the metallicity profiles. In Fig.~\ref{fig:ageM}, we show the age gradients as a function of the stellar mass of the simulated galaxies. On average, the mean age gradients are negative as expected for inside-out histories of formation. The  slopes of the age profiles show a very mild dependence on stellar mass.  In Table~\ref{table:1} we show the median age gradients for the two stellar-mass subsamples. The normalized age slopes show a similar behaviour, again with a large dispersion. The Spearman test show very weak signal for the non-normalized relations ($r =0.19, p= 0.26$; upper panel) and no signal at all for the normalized one ($r =-0.044, p=0.80$; lower panel).

As a function of $r_{\rm eff}$ the age gradients show  no clear signal of correlation. This might be caused by the intrinsic dispersion in stellar age or by the low number of galaxies. In  analytical models, a parallel behaviour between metallicity and age could be expected but when the non-linear assembly of structure is considered, then the relations are {more complex}. And in fact the age profiles could be more affected by fraction of young stars as a function of radius. In our simulation as we will see later on, there are high metallicity stars of all ages, although on average, they tend to be younger compared to low-metallicity ones.

We note that there are steeper negative normalized age gradients in the simulation than in the observations reported by \citet{sanchezb2014}. These age profiles with steeper negative slopes could be the result of a difference in the estimation of the $r_{\rm eff}$, and/or  of the action of  physical processes  poorly described in our model or differences in the galaxy sample. 

The inside-out disc formation scenario seem to be the common  formation path for our discs  in agreement with observations. However, there are few of them which show positive age profiles (less than ten per cent). We found that these discs have  higher fractions of  recent new-born stars within  $r_{\rm eff}$ which are responsible for the flattening or changing in the slope of the age profiles as expected. However, we found that not all of them show positive metallicity slopes. In previous works, the positive metallicity slopes of the gas-phase components were shown to appear as a result of the effects of  mergers or interactions which drove gas infalls triggering star formation. These gas inflows are expected to have lower metallicity than the gas present in the central regions \citep[e.g.][]{perez2006,rupke2010,perez2011,tissera2015}. However, if these inflows took place very recently, the gas-phase disc would be already enriched \citep{tissera2015}. So the new-born stars would be young but not so metal-poor with respect to the underlying SPs. Hence the age profiles would reflect the recent star formation activity  in the central regions more efficiently than the the metallicity gradients. Our simulations suggest that there might be a variety of situations, depending on the metallicity of the infalling gas (i.e. the slope of the gas-phase metallicity gradients) and the relative fraction of young-to-old stars present in the central regions.

\begin{figure*}
\includegraphics[width=0.85\columnwidth]{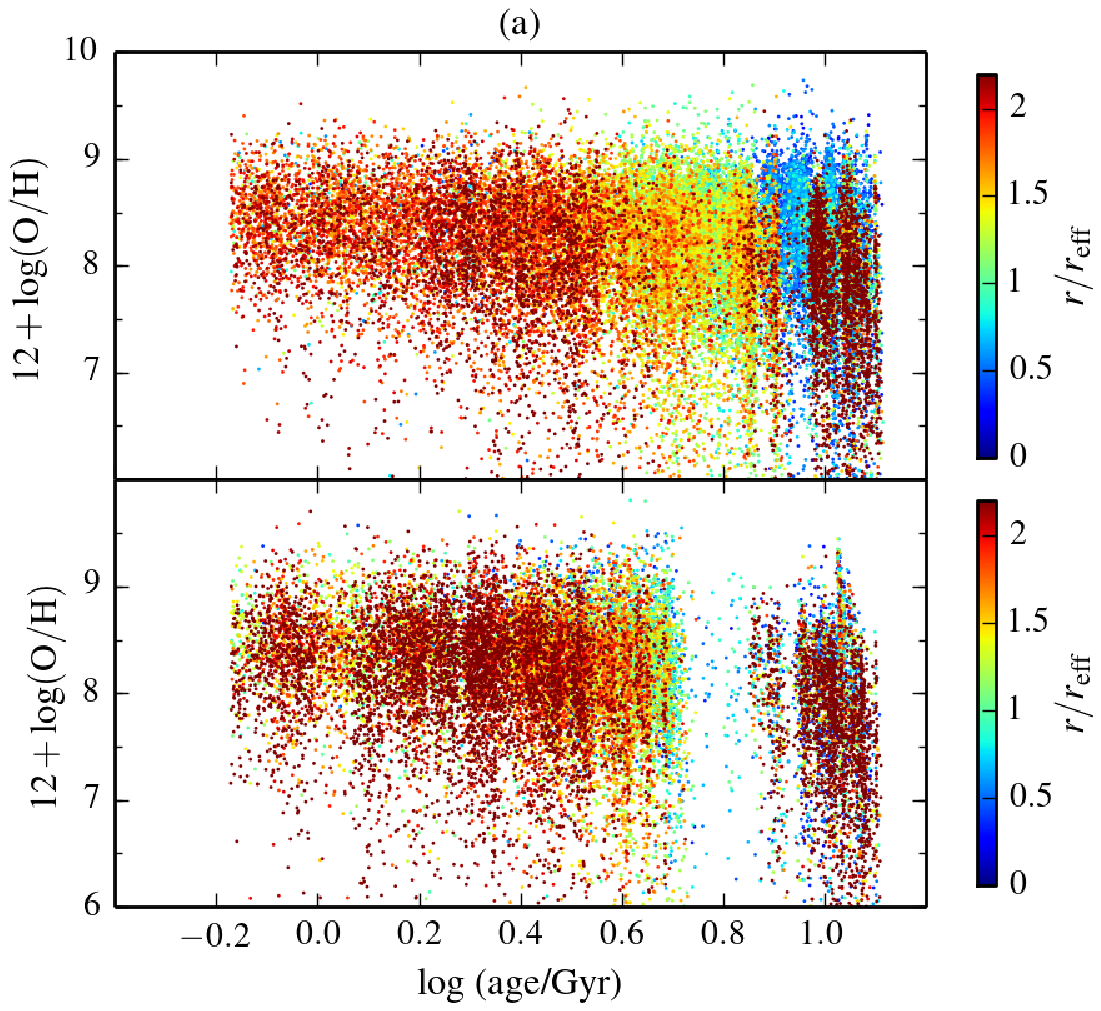}
\includegraphics[width=0.85\columnwidth]{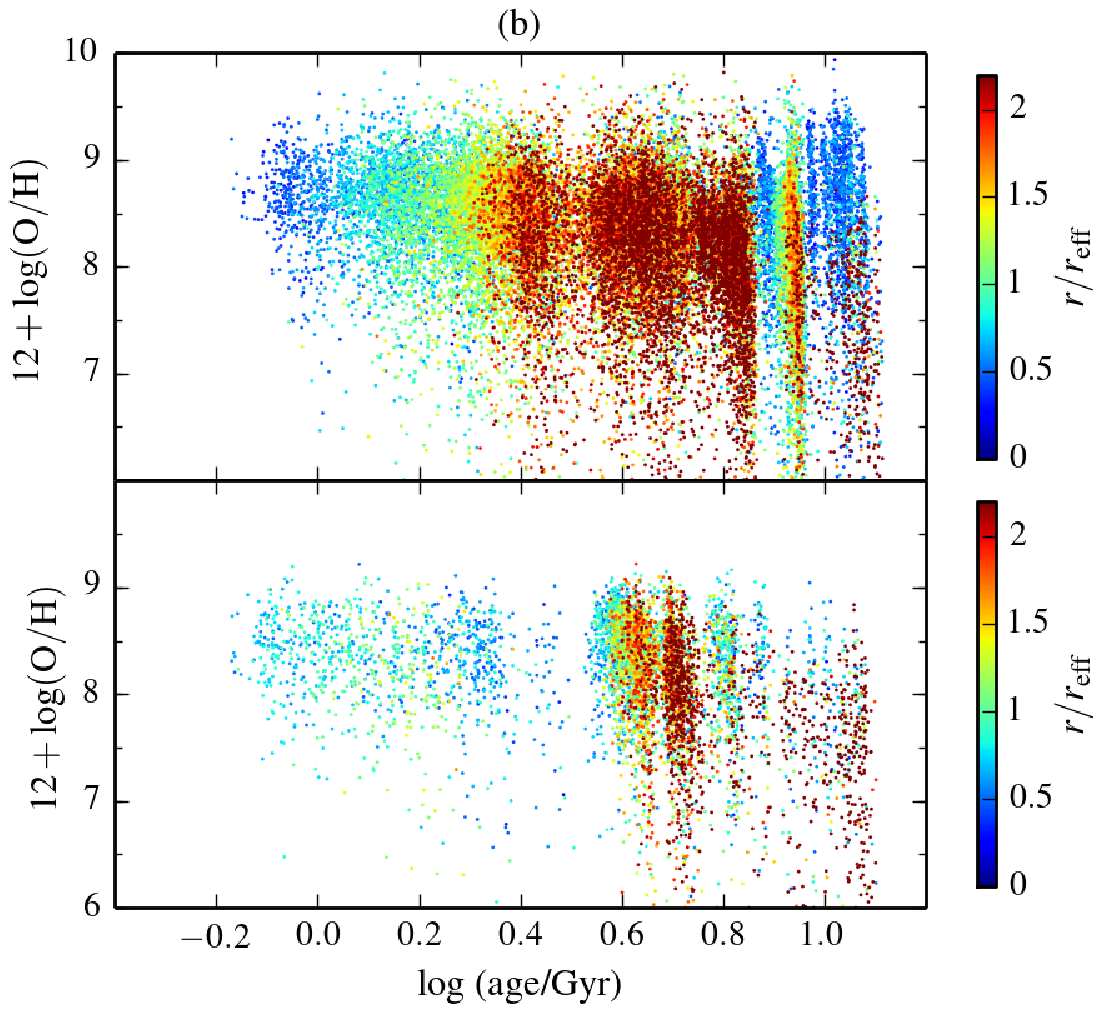}
\centering
\caption{Age-metallicity relation for simulated galaxies: (a)
   negative age gradients  (upper-left galaxy: $\nabla_{\rm age}=-0.41;\nabla_{\rm Z} = -0.34$; lower-left
   galaxy:$\nabla_{\rm age}=-0.40; \nabla_{\rm Z}=-0.14$)  and (b)  positive age
  gradients (upper-right galaxy: $\nabla_{\rm age}= 0.32;
  \nabla_{\rm Z}=-0.34$; lower-right galaxy:$\nabla_{\rm age}= 0.10;  \nabla_{\rm Z}=-0.14$).
The colours denote the distance  to the centre of mass in units of the
$r_{\rm eff}$. Each point depicts a star particle representing a
single stellar population.
The relation between old and new
SPs is clearly different and is the responsible for the different age
profiles in the simulations.}
\label{fig:agemet}
\end{figure*}

For illustration in  Fig.~\ref{fig:agemet}, we show the age-metallicity relation for two galaxy discs with positive (right panel)  and negative (left panel) age gradients, which  have negative metallicity slopes.  From this figure we can see that the galaxies with  positive age slopes have a larger fraction of young stars  within $r_{\rm eff}$.  Those with negative age slopes have  young stars formed mainly in the outer parts of the discs or more evenly distributed along the discs. We expect positive age slopes if there is available gas to form stars in the central regions, and this might happen if  the gas is  transported inward by tidal torques produced during interactions with nearby satellites or galaxies.

We also find that old stars are present at all radii. In particular, in Fig.~\ref{fig:agemet}  the presence of old stars in the outer regions can be clearly seen. This could be ascribed in part  to migration processes \citep[e.g.][]{spitoni2015} or bar diffusion \citep[][]{brunetti2011}. It  should be also mentioned that the accretion of satellite galaxies could also contribute with old stars in the outer parts \citep{abadi2003,tissera2012}. 

Hence, it might be possible to have galaxies with positive age profiles which have experienced secular evolution recently but their metallicity gradients still show negative slopes. This situation adds noise to observations and may prevent  drawing clear conclusions in relation to  the presence of bars and their impact on metallicity slopes \citep{sanchezb2014}. 

\subsection{Metallicity gradients for old and young stars}

\begin{figure}
\includegraphics[width=0.8\columnwidth]{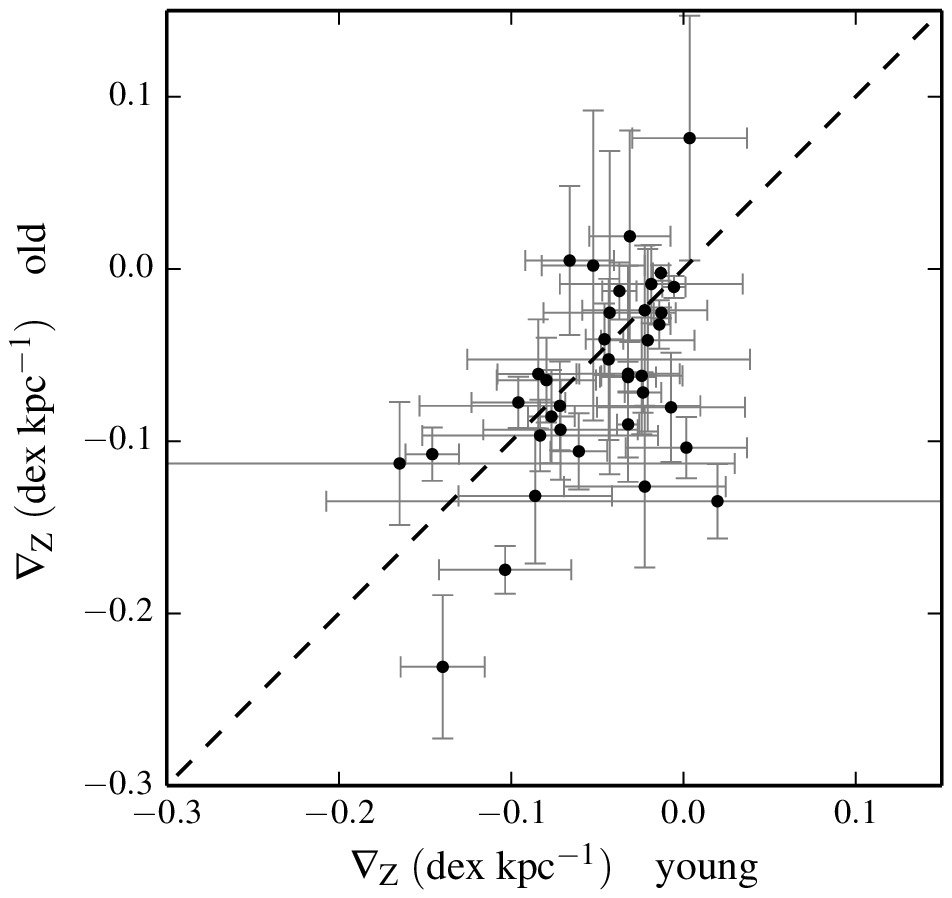}
\centering
\caption[]{Slopes of the normalized metallicity profiles  of the old stellar population ($>
  6$~Gyr) versus those of the young stellar population ($< 2$~Gyr) for
  the simulated disc components.}
\label{fig:oldyoung}
\end{figure}

\citet{sanchezb2014} reported a trend, albeit weak, for old stellar population to have normalized metallicity profiles with steeper negative slopes than those determined by young stars. To evaluate this aspect in our simulated galaxies, we estimated the metallicity profiles for stars older than 6~Gyr and younger than 2~Gyr for both normalized and non-normalized profiles. Figure ~\ref{fig:oldyoung} shows the estimated metallicity gradients. The median values have a mild trend for the old populations to have steeper metallicity gradients ($-0.06 \pm 0.01$ dex~ kpc$^{-1} $) compared to the young ones ($-0.04 \pm 0.02$ dex~kpc$^{-1} $). A stronger trend is found for the normalized metallicity slopes although with   larger bootstrap errors and dispersions (Table~\ref{table:2}). 

\begin{figure}
\includegraphics[width=\columnwidth]{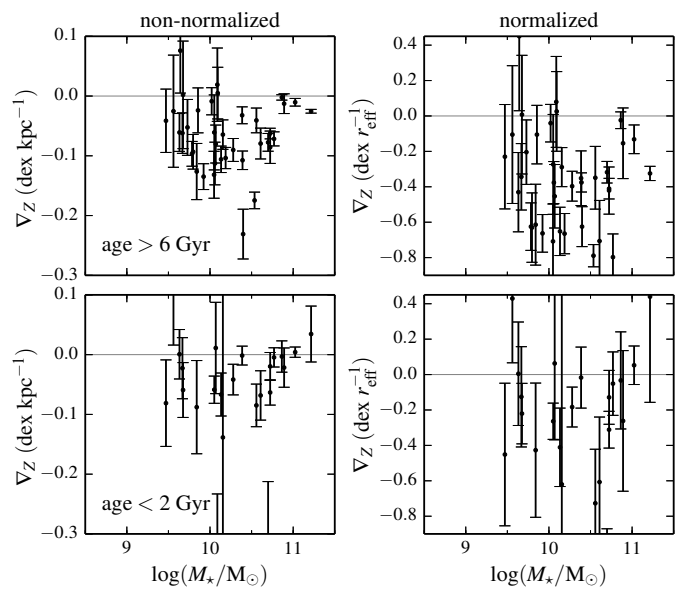}
\centering
\caption[]{Normalized (left panels)  and non-normalized (right panels)
  metallicity slopes  as a function of stellar mass separated into two
  subpopulations: old ( $>$~6 Gyr; upper panels) and young  $<$~2 Gyr;
  lower panels).}
\label{fig:ages}
\end{figure}

For both age subsamples, we estimated the slopes of the stellar metallicity profiles as a function of the galaxy stellar mass as shown in Fig.~\ref{fig:ages}.  The non-normalized gradients show a very weak correlation with stellar mass for the young SPs The  slopes of the normalized metallicity profiles show no correlations for the old and young SPs with stellar mass   (Table \ref{table:2}). The  normalization of the metallicity profiles by the corresponding $r_{\rm eff}$ erases any dependence on the stellar mass that might have been imprinted on the metallicity profiles of the young stellar population. Note that the Spearman tests provide no clear signal of correlations. Hence, the trends estimated above should be taken as indicative and be confirmed with a larger sample. We note that both the non-normalized and the normalized  metallicity gradients of the old SPs show  a clear U-shape distribution (the young populations are noiser). From these trends it seems that the  old populations are the primarly responsible for this behaviour in the simulation (Fig~\ref{fig:metM} and Fig.~\ref{fig:metR2}).
 
The gas-phase metallicity profiles show a clearer dependence on stellar mass as reported by \citet{tissera2015}. However the SPs are not able to  preserve this correlation in our simulations and even for the younger stars, the signal is very weak. This suggests the action of mechanisms which contribute to mix up stars such as mergers/interactions and/or migration as previously mentioned.

\begin{table*}
\caption{Median metallicity ($\nabla_{Z}$) and age($\nabla_{\rm age}$)  gradients
  for the normalized (${\rm dex~{\it
      r}_{eff}^{-1}}$) and non-normalized  (${\rm dex~kpc^{-1}}$)
  stellar profiles.
   The medians as well as the
  errors are estimated via a bootstrapping technique (the  standard
  deviations are given  within parentheses).} 
\label{table:1}
\centering
\begin{tabular}{l c c c c c}
\hline\hline
& $M_{\star}$         & $\nabla_{Z}$   & $\nabla_{Z}$   & $\nabla_{\rm age}$   & $\nabla_{\rm age}$\\
& $({\rm M}_{\odot})$ & $({\rm dex~kpc^{-1}})$ & $({\rm dex~{\it r}_{eff}^{-1}})$ & $({\rm dex~kpc^{-1}})$ & $({\rm dex~{\it r}_{eff}^{-1}})$\\ 
\hline
\multirow{2}{*}{Simulation} & $ < 10^{10.5} $ & $-0.04 \pm 0.01~(0.04)$ & $-0.21 \pm 0.07~(0.18)$ & $-0.04 \pm 0.01~(0.04)$ & $-0.17 \pm 0.06~(0.20)$ \\
                            & $ > 10^{10.5} $ & $-0.03 \pm 0.02~(0.05)$ & $-0.26 \pm 0.10~(0.27)$ & $-0.02 \pm 0.01~(0.02)$ & $-0.23 \pm 0.12~(0.19)$ \\
\\
\hline
\end{tabular}
\end{table*}

\begin{table*}
\caption{Median metallicity gradients, both normalized to the
  effective radius of each disc (in ${\rm dex~{\it r}_{eff}^{-1}}$)
  and non-normalized (in ${\rm dex~kpc^{-1}}$). Galaxies are separated
  into two subsamples according to  stellar mass, and the gradients were measured for the old and the young stellar populations separately. Errors are as in Table~\ref{table:1}.} 
\label{table:2}
\centering
\begin{tabular}{l c c c}
\hline\hline
& $M_{\star}$         & $\nabla_{Z}$   & $\nabla_{Z}$ \\
& $({\rm M}_{\odot})$ & $({\rm dex~kpc^{-1}})$ & $({\rm dex~{\it r}_{eff}^{-1}})$  \\ 
\hline
\multirow{3}{*}{Age $>$ 6 Gyr} & $ < 10^{10.5} $ & $-0.07 \pm 0.01~(0.06)$ & $-0.36 \pm 0.06~(0.28)$  \\
                               & $ > 10^{10.5} $ & $-0.06 \pm 0.02~(0.05)$ & $-0.37 \pm 0.10~(0.25)$  \\
                               & All masses      & $-0.06 \pm 0.01~(0.06)$ & $-0.36 \pm 0.04~(0.28)$  \\
\\
\multirow{3}{*}{Age $<$ 2 Gyr} & $ < 10^{10.5} $ &  $-0.05 \pm 0.02~(0.09)$ & $-0.23 \pm 0.11~(0.39)$   \\
                               & $ > 10^{10.5} $ &  $-0.03 \pm 0.02~(0.09)$ & $-0.21 \pm 0.16~(0.46)$   \\
                               & All masses      &  $-0.04 \pm 0.02~(0.09)$ & $-0.22 \pm 0.09~(0.42)$  \\
\hline
\end{tabular}
\end{table*}

\section{Conclusions}
\label{sec:conclu}

We studied the metallicity and age gradients of the stellar discs  $z \sim 0$  in a hierarchical clustering scenario with the aim of analysing if they reproduce observational trends and how they depend on stellar mass. The metallicity gradients reflect the enrichment of the ISM at the time of star formation and hence, can store information which is relevant to understand the star formation process, the enrichment cycle and the effects of galaxy assembly.

Our results can be summarized as follows:

\begin{itemize}

\item The non-normalized  metallicity gradients of the SPs show a very weak dependence on stellar mass of the galaxies which is erased  when normalized by  $r_{\rm eff}$. There is a large variety of normalized metallicity gradients. The results are in global agreement with those  reported by CALIFA survey \citep{sanchezb2014}, considering the differences in the definition of the scale-lengths (Section 3). Our simulated discs show the steeper metallicity gradients for galaxies with stellar masses around $10^{10}$M$_\odot$ in agreement with observational results \citep[e.g.][]{gonzalezdelgado2015}.

\item The simulated  $r_{\rm eff}$ of the stellar discs as a function of stellar mass are consistent with observational estimations by \citet{wel2014} and by \citep{sanchezb2014}, within a standard deviation. Our simulated stellar discs formed by conserving the specific angular momentum content \citep{pedrosa2015} with mean values in agrement with observations \citep{romanowsky2012}. These are important achievements for the our simulations considering that the free parameters were not fixed to reproduce them.

\item We also find a correlation between the metallicity gradients and the $r_{\rm eff}$ which is erased if the profiles are normalized by this characteristic scale-length. As pointed out by previous works \citep{prantzos2000}, this suggests that the building up of the metallicity profiles is strongly linked with the formation of discs. Other processes such as galaxy mergers, secular evolution, migration, accretion of small satellites could perturbate the discs and mix the gas and stellar populations producing changes in the relation. As reported in previous works, we expect mergers to be the one of the  most important mechanisms responsible for driving  inflows which contribute to produce positive metallicity and/or age profiles \citep{rupke2010,perez2006,perez2011}. In fact, the metallicity gradients may change back to negative but  they tend to remain with values above the original ones.  Also, the ejection of enriched material from the central regions by SN outflows which could be re-accreted in the external regions of the discs later on, and contribute to flatten the metallicity profiles \citep{perez2011}.

\item The simulated stellar discs show negative age profiles consistent with an inside-out history of formation, on average. There are some discs with positive (or close to flat) age gradients which  can be linked to the recent formation of  stars in the central region $ r/r_{\rm  eff} < 1$. These cases could be related to galaxies with recent history of galaxy  interactions which have driven gas inflows. The metallicity profiles are weakly modified since recent gas infalls would tend to transport already enriched material. Conversely the age profiles are strongly modified by these new stars since the dynamical range covered by age is much larger.
Our results suggest that, in some cases,  relating the presence of a bar as responsible for gas inflow, with the slope of metallicity profile might not be so direct.
 
\item The stellar populations show metallicity profiles with different slopes when they are separated according to age. The metallicity gradients of young ($< 2$~Gyr) stars tend to be slightly flatter than those of the  old ($> 6$~Gyrs) stars. These trends are in agreement with results from {\small CALIFA} survey and the Milky Way \citep{maciel2003}.

\end{itemize}

Overall, our findings suggest that the underlying process responsible for determining  the age and metallicity profiles  is the inside-out formation of disc, with specific angular momentum conservation. In spite of this, dynamical processes which can disturbe the stellar distributions together with those that affect the transformation of gas into stars play an important role. 

\begin{acknowledgements}
Support for PBT is provided by the Ministry of Economy, Development, and Toursim's Millennium Science Initiative through grant IC120009, awarded to The Millennium Institute of Astrophysics, MAS. REGM acknowledges support from \textit{Ci\^encia sem Fronteiras} (CNPq, Brazil). This work was partially supported by PICT 2011-0959 (ANCyT, Argentina) and PIP 2012-0396 (Conicet, Argentina). PBT acknowledges partial support from  the Regular Grant UNAB 2014, Nucleo UNAB 2015 (DI-677-15/N) of Universidad Andres Bello and Fondecyt 1150334 (Conicyt).
\end{acknowledgements}

\bibliographystyle{aa}
\bibliography{SPDisc_rv2}

\end{document}